\documentclass[prd]{revtex4}
\usepackage{graphicx,amsmath,amsfonts}

\newcommand\nn{\nonumber}
\newcommand\ba{\begin{eqnarray}}
\newcommand\ea{\end{eqnarray}}
\newcommand{\br}[1]{\left( #1 \right)}
\newcommand{\brs}[1]{\left[ #1 \right]}

\begin{document}

\title{Radiative corrections in  $K_{e4}$ decay}
\author{Yu.~M.~Bystritskiy}
\email{bystr@theor.jinr.ru}
\affiliation{Joint Institute for Nuclear Research, 141980 Dubna,Russia}

\author{S.~R.~Gevorkyan\footnote{On leave of absence from Yerevan Physics Institute}}
\email{gevs@jinr.ru}
\affiliation{Joint Institute for Nuclear Research, 141980 Dubna,Russia}

\author{E.~A.~Kuraev}
\email{kuraev@theor.jinr.ru}
\affiliation{Joint Institute for Nuclear Research, 141980 Dubna,Russia}

\begin{abstract}
The final state interaction of pions in the decay   $K^\pm\to \pi^+\pi^-e^\pm\nu $
 allows   to obtain  the value of the isospin  and angular momentum  zero  pion-pion
scattering length $a_0^0$.  To extract this quantity from  experimental data
the radiative corrections (RC) have to  be taken into account.
Basing on the lowest order results  and  the factorization hypothesis,
we get  the expressions  for RC in the leading and next-to leading logarithmical approximation.
It is shown that the decay width dependence on the lepton mass $m_e$  through  the parameter
 $\sigma=\frac{\alpha}{2\pi}\br{\ln\frac{M^2}{m_e^2}-1}$   has  a standard form
of the Drell-Yan process and is  proportional to the Sommerfeld-Sakharov factor.
The numerical estimations are presented.
\end{abstract}

\maketitle

\section{Introduction}

Kaons decay with two or three pions in the final state could give the unique information
on the value of the $s$ and $p$-wave pion-pion  scattering lengths, whose values are
predicted very precisely within Chiral Perturbation Theory~\cite{Colangelo:2001df}.
The semi-leptonic decay
known as  $K_{e4}$ decay (for definiteness we will discuss  the $K^+$ decay)
(see Fig.~\ref{FigBorn}):
\begin{figure}
\begin{center}
\includegraphics[width=0.4\textwidth]{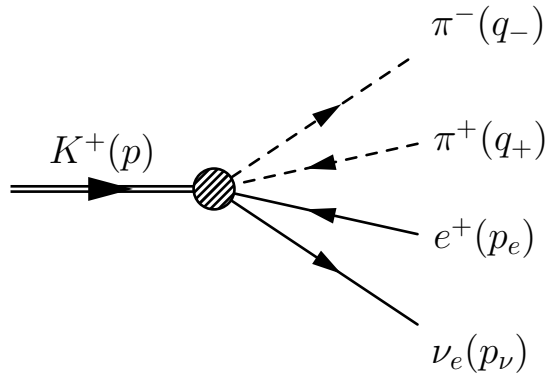}
\caption{The semi-leptonic decay $K_{e4}$.
\label{FigBorn}}
\end{center}
\end{figure}
\ba
 K^\pm(p)\to\pi^+(q_+)+\pi^-(q_-)+e^\pm(p_e)+\nu(p_\nu)
 \label{Born}
\ea
 is  very clean environment for the measurement of $\pi\pi$ scattering
lengths, since the two pions  are the only hadrons in the final state  produced close to threshold.\\
Recently the high statistics measurement of  $K_{e4}$ decay has  been done by NA48/2 collaboration at the
CERN SPS~\cite{Batley:2007zz}.The high quality of this data allows one to extract the scattering length $a_0^0$
with accuracy comparable with theoretical predictions. From the other hand to obtain such high precision
in scattering length determination from experimental data one would take into account  all effects,
which can have impact on the value of extracting quantity. One of   such effects crucial
 in obtaining the scattering length value from experimental data  is the correct accounting
 of radiative corrections  in the decay (\ref{Born}). For  RC calculations in decays like  the Monte Carlo package
 PHOTOS has been developed ~\cite{Barberio:1990ms,Barberio:1993qi} and widely used in data processing. Unfortunately the
 PHOTOS  does not take into account the electromagnetic interaction between charged pions in the final state effect, which is
  important near production threshold~\cite{Gevorkyan:2006rh} when the relative velocity in the pion pair becomes small.
  Moreover the Monte Carlo calculations are not transparent and require the special consideration of accuracy for any
  particular decay.\\
   From the other hand a large improvement in accounting the  RC  in the decays with several hadrons in final state
 has been done recently~\cite{Isidori:2007zt, Bissegger:2008ff}. Later on we   consider the full set of   RC  in the $K_{e4}$
 decay and obtain  the relevant expressions, which can be easily applied to actual calculations in the decay (\ref{Born}).
  Our expressions are in accordance with ones from~\cite{Isidori:2007zt}  with two main difference. Despite the expressions
 in ~\cite{Isidori:2007zt} our formulas  are suitable for the case when one of the particle in the  final state (lepton) differ from other (pions).
  Moreover we take into account also the radiation of hard photons, which leads to disappearance of  cut in photon energy.\\
Before  consideration of the proper RC  let us shortly discuss the widely exploited approach  ~\cite{Pais:1968zz}
 to  the process (\ref{Born})  without  electromagnetic effects.The relevant  matrix element can be expressed as
\ba
T=\frac{G_F}{\sqrt{2}}V_{us}^*\left(V^\mu-A^\mu\right)\bar{u}(p_\nu)\gamma_\mu(1-\gamma_5)v(p_e),
\label{BornAmplitude}
\ea
where the axial and vector hadronic currents
\ba
A^\mu&=&\frac{-i}{M}\br{\br{q_++q_-}^\mu F+\br{q_+-q_-}^\mu G+\br{p_e+p_\nu}^\mu R}; \nn\\\\
V^\mu&=&\frac{-H}{M^3}\epsilon^{\mu\nu\rho\sigma}p_\nu\br{q_++q_-}_\rho \br{q_+-q_-}_\sigma,
\ea
where $M$ is the K-meson mass.
The contribution of the axial form factor  R  to the differential width is proportional to
the square of electron mass  and  would be omitted.
Confining   by  s and p waves  and assuming the same p-wave phases for different  form-factors:
\ba
F=F_se^{i\delta_s}+F_pe^{i\delta_p};
\qquad
G=G_pe^{i\delta_p};
\qquad
H=H_pe^{i\delta_p}.
\ea
The aim of experimental investigation is to measure the quantities  $F_s$, $F_p$, $G_p$, $H_p$  and the phases difference  $\delta=\delta_s-\delta_p$
as a function of dimensionless invariants $s_\pi=(q_++q_-)^2$, $s_e=(p_\nu+p_e)^2$.

Besides these  variables there are  three angles  in  use. Azimuthal angle $\phi$ between the plane containing the
pions  momenta in the kaon  rest frame  and the plane containing the electron and neutrino
momenta;  the polar angle $\theta_\pi$ between the positive charged pion and the dipion
line  and finally the polar angle $\theta_e$ between the electron momentum and the dilepton  line.\\
The differential width has  the form ~\cite{Pais:1968zz}
\ba
d\Gamma_B=\frac{G_F^2|V_{us}|^2}{2(4\pi)^6}\Lambda^{1/2}(M^2,s_\pi,s_e)
\beta(1-\frac{m_e^2}{s_e})^2 J ds_\pi ds_ed\cos\theta_\pi d\cos\theta_e d\phi,
\label{DiffWidth}
\ea
where
\ba
\Lambda(a,b,c) &=&a^2+b^2+c^2-2(ab+ac+bc)\nonumber
\ea
The structure $J=J(s_\pi,s_e,\theta_\pi,\theta_e,\phi)$
is the rather complicate function of four form-factors~ \cite{Pais:1968zz}, whereas
$\beta=\sqrt{1-\frac{4m^2}{s_\pi}}$ (m is the charged pion mass)
is the relative velocity of pions in the kaon rest frame.

Calculation of radiative corrections , which is the motivation of our paper is performed
in frames of unrenormalized theory.  We introduce the fictitious mass of the photon
$\lambda$ and momentum cut-off parameter $\Lambda$. The final result which takes into account
emission of virtual  and real photons would  be free from infrared divergences connected with
photon mass. Keeping in mind the renormalizability of the Standard Model
the cut-off parameter $\Lambda$
at the final stage  must be replaced by the W  boson mass $M_W$.

Our paper is organized as follows. The explicit calculation of contributions of channels with virtual
and real (soft and hard) photons in lowest  order in fine structure constant  are presented in the first two sections.
 The combined result in the lowest order of perturbation theory and its generalization to higher orders  are
 given in two following sections.

Appendix~\ref{AppendixIntegrals} contains the details of calculations of virtual
and real photons emissions.
Appendix~\ref{AppendixKFactors} contains the explicit forms  of
$K$, $K_v$, $K_s$  factors using in numerical calculations.

In Table~\ref{TableKFactors} the result of a numerical estimations of a
width and  $K$  factors  are given for several typical values
of the kinematical invariants.

\section{Virtual photons emission}

\begin{figure}
\begin{center}
\includegraphics[width=0.8\textwidth]{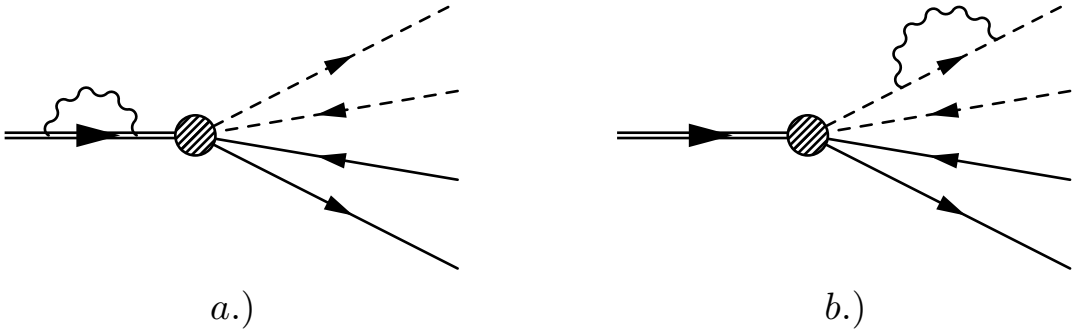}
\caption{Some typical graphs responsible for the
renormalization of the wave functions.
\label{FigWF}}
\end{center}
\end{figure}

\begin{figure}
\begin{center}
\includegraphics[width=0.8\textwidth]{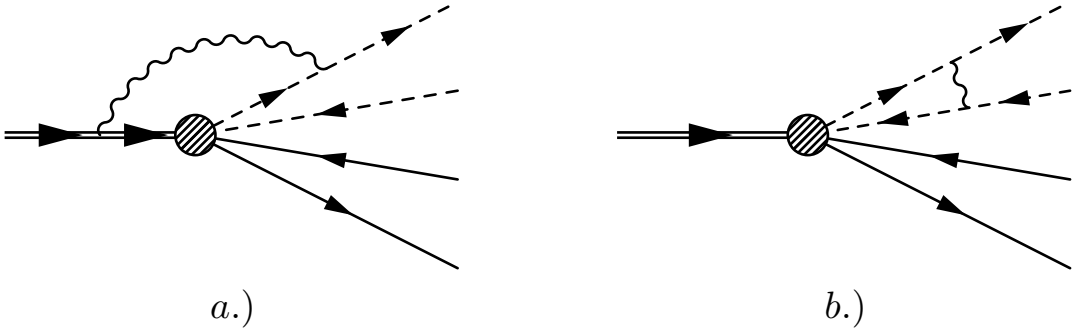}
\caption{Some graphs describing the virtual corrections to $K_{e4}$ decay.
\label{FigVirt}}
\end{center}
\end{figure}

Let us at first  shortly  discuss the corrections  arising from the virtual photon emission.
An important ingredient in such consideration  is the wave functions renormalization constants
of electron and pseudoscalar mesons (see Fig.~\ref{FigWF})
\ba
Z_e &=&\frac{\alpha}{2\pi}\brs{-\frac{1}{2}L_\Lambda-\frac{3}{2}L_e+L_\lambda-\frac{9}{4}};
\quad
Z_P=\frac{\alpha}{2\pi}\brs{L_\Lambda+L_\lambda-\frac{3}{4}}
\nn\\   L_\Lambda &=&\ln\frac{\Lambda^2}{m^2},
\quad
L_e=\ln\frac{m^2}{m_e^2},
\quad
L_\lambda=\ln\frac{m^2}{\lambda^2}.
\ea
The relevant contribution to the differential widths (\ref{DiffWidth}) can
be introduced by replacement
\ba
J\to J\br{1+3Z_p+Z_e}
\ea
Neglecting structure emission (when photons  are  emitted from "hard" hadronic or weak blocks)
we have to  consider six Feynman  amplitudes with virtual photon attached to  charged particles
(see Fig.~\ref{FigVirt}).
Neglecting as well by the virtual photon momentum in the "hard" block we obtain
\ba
T^v&=&\frac{G_F V_{us}^*}{\sqrt{2}}(V^\mu-A^\mu)\frac{\alpha}{4\pi}
\int\frac{d^4k}{i\pi^2(k^2-\lambda^2)}[\sum_{i=1}^3 R_i \bar{u}(p_\nu)\gamma_\mu(1-\gamma_5)v(p_e)\nn\\
&+& \sum_{i=1}^3Q_i^\eta\bar{u}(p_\nu)\gamma_\mu(1-\gamma_5)
(-\hat{p}_e-\hat{k}+m_e)\gamma_\eta v(p_e)],
\label{VirtAmplitude}
\ea
with the following notations
\ba
R_1&=&\frac{(-2q_+-k)_\sigma(-2p-k)^\sigma}{d d_+};
\quad
R_2=\frac{(2q_-+k)_\sigma(-2p-k)^\sigma}{d d_-};\nn\\
R_3&=&\frac{(-2q_++k)_\sigma(2q_-+k)^\sigma}{d_-d_+}, \nn\\
Q_1^\eta&=&\frac{(-2p-k)^\eta}{d d_e}; \quad  Q_2^\eta=\frac{(-2q_+k)^\eta}{d_+d_e};
\quad Q_3^\eta=\frac{(2q_--k)^\eta}{d_-d_e},\nn\\
d&=&(p+k)^2-M^2+i0;\quad d_+=(-q_++k)^2-m^2+i0;\nn\\
  d_-&=&(q_--k)^2-m^2+i0;\quad d_e=(p_e+k)^2-m_e^2+i0.
\ea
The contribution of virtual photon loops to the decay rate (\ref{DiffWidth}) is determined by
the real part of the interference
between the single loop and the Born amplitude (\ref{BornAmplitude}).
The standard integration of expression (\ref{VirtAmplitude})  leads to the following  form
of this interference
\ba
J\br{1+3Z_p+Z_e+\frac{\alpha}{2\pi}\brs{I_1+I_2+I_4+I_3+I_5+I_6}}.
\label{JDef}
\ea
The explicit form of the six integrals  $I_i$ are  given
in Appendix~\ref{AppendixIntegrals}.
The assumption about smooth behavior of the structure
$J(s_\pi,s_l,...)=J_0$ allow us to write down the
contribution from the emission of virtual photons as
\ba
\frac{d\Gamma^v}{d\Gamma_B}&=&\frac{\alpha}{2\pi}
\left[
    L_\lambda\br{4+\frac{1}{\beta_-}L_--\frac{1}{\beta_+}L_+
    -2\rho-\frac{1+\beta^2}{\beta}L_\beta+2\ln{\frac{p_eq_+}{p_eq_-}}}
\right.
\nn\\
&+&
\left.
\pi^2\frac{1+\beta^2}{\beta}-2l\rho+4\rho+\frac{1}{2}\ln^2\frac{M^2}{m_e^2}+\frac{9}{4}L_\Lambda+K_v
\right],
\label{VirtRC}
\ea
Here $\rho=\ln\frac{2E_e}{m_e}$ is the
"large logarithm" ($E_e$, $E_\pm$ is the positron and
pions energies in the kaon rest frame)
\ba
L_\pm&=&\ln\frac{1+\beta_\pm}{1-\beta_\pm}; \qquad \beta_\pm=\sqrt{1-\frac{m^2}{E_\pm^2}}\nn\\
l&=&\ln\frac{M^2}{m^2};\qquad  L_\beta=\ln\frac{1+\beta}{1-\beta};
\ea
The explicit form of  $K_v$  is  cited  in Appendix~\ref{AppendixKFactors}.

\section{Real photons emission}

Let us now discuss the emission of real photons.
The contribution of soft (in the kaon rest frame) photons
is proportional to the decay  width in Born
approximation (see Fig.~\ref{FigSoft}):
\begin{figure}
\begin{center}
\includegraphics[width=0.8\textwidth]{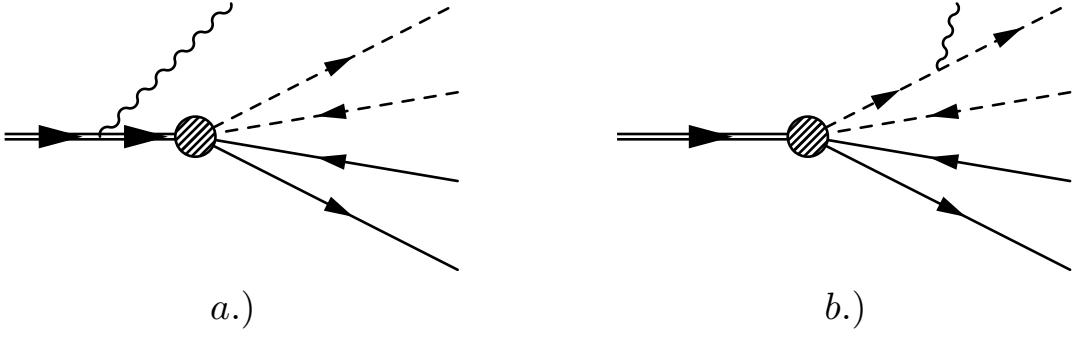}
\caption{Real photon emission corrections to $K_{e4}$ decay.
\label{FigSoft}}
\end{center}
\end{figure}
\ba
\frac{d\Gamma^{soft}}{d\Gamma_B}&=&
-\frac{\alpha}{4\pi^2}\int\frac{d^3k}{\omega}
\br{\frac{p}{pk}+\frac{q_-}{q_-k}-\frac{q_+}{q_+k}-\frac{p_e}{p_ek}}^2_{\omega<\Delta\epsilon}\nn\\
&=&\frac{\alpha}{\pi}\Delta^s, \qquad\Delta\epsilon \ll E_e.
\label{SoftIntegration}
\ea
The standard calculations  give
\ba
\Delta^s&=&\ln\br{\frac{2\Delta\epsilon}{\lambda}}
\brs{-4-\frac{1}{\beta_-}L_-+\frac{1}{\beta_+}L_++2\rho+
\frac{1+\beta^2}{\beta}L_\beta-2\ln\frac{2p_eq_+}{2p_eq_-}}\nn\\
&+&\rho-\rho^2+ K_s.
\ea
with  expression for $K_s$ given in Appendix \ref{AppendixKFactors}.

It is easy to see that the  sum of soft
(eq. (\ref{SoftIntegration})) and virtual (eq.(\ref{VirtRC})) photons does not depend
on the introduced above fictitious  photon  mass $\lambda$.

At small relative velocity of pions $\beta$ the
term $\frac{\pi\alpha(1+\beta^2)}{2\beta}$ in (\ref{VirtRC}) corresponds to the well
known Sommerfeld-Sakharov factor \cite{Sommerfeld:1939,Sakharov:1948yq}
\ba
S(\beta)=\frac{t}{1-exp(-t)}=1+\frac{1}{2}t+\frac{t^2}{12}+O(t^3),
\qquad
t=\frac{\pi\alpha(1+\beta^2)}{\beta}.
\ea
Due to the general statements of quantum mechanics this factor is factorized out from the
differential width for the case of small $\beta$.\\
All  terms containing the positron mass singularities (which contains the quantity $\rho$)
can be written in form of the so called delta-part of
positron non-singlet structure function $P_\delta$.
As a result the contribution of soft and virtual photons  can be written as
\ba
1&+&\frac{d\Gamma^s+d\Gamma^v}{d\Gamma_B}=\brs{1+\sigma P_\delta}
\br{1+\frac{\alpha}{\pi}K}~S(\beta)~S_{EW}, \\
P_\delta&=&2\ln\Delta+\frac{3}{2},  \quad
\sigma=\frac{\alpha}{2\pi}(2\rho-1), \quad
S_{EW}=1+\frac{9\alpha}{4\pi}\ln\frac{M_W^2}{m^2}, \nn
\ea
where $\Delta=\frac{\Delta\epsilon}{E_e} \ll 1$.

The factor $S_{EW}$ is absorbed, when we use the renormalized quantities instead
of bare ones
\ba
(G_F^2V_{us}^2)^{bare}S_{EW}=G_F^2V_{us}^2.
\ea
The expression for the quantity $K$ given in Appendix~\ref{AppendixKFactors}.
The values of  $K$, $K_v$, $K_s$
for several typical sets of the kinematic parameters are tabulated in Table~\ref{TableKFactors}.

It is convenient to separate the contribution from the emission of hard photons
$\omega>\Delta\epsilon$
in two parts. First one takes into account the emission along the positron direction.
Another one takes into account the remaining part of the angular phase volume.

The first one can be calculated using the so called
"quasi-real electrons" method \cite{Baier:1973ms}:
\ba
d\Gamma^h(s_\pi,s_l,...)=
\int\limits_{s_l(1+\Delta)}^{s_{max}}\frac{d s}{s}
\brs{P_\theta(\frac{s_l}{s})\sigma+
\frac{\alpha}{2\pi}(1-\frac{s_l}{s})}d\Gamma_B(s_\pi,s,...),
\ea
with $s_{max}=(p-q_+-q_-)^2=M^2+s_\pi-2M(E_++E_-)$
\ba
P_\theta(z)=\frac{1+z^2}{1-z}.
\ea
As for the contributions which is not enhanced by the "large logarithm" factor  their
contribution can be  estimated in the soft photon emission approximation. It can be
obtained from the quantity $\Delta^s$ putting $\Delta\epsilon=\omega_0=M-2m$.
Soft photons approximation turns out to be rather realistic. The typical error compared with
the exact calculation in the same order of perturbation theory looks as
\ba
1+O\br{\br{\frac{\omega}{M}}^2},
\quad
\omega<\omega_{max}=M-2m,
\quad
\br{\frac{\omega}{M}}^2\sim 0.1.
\ea
Combining the Born approximation and the lowest order results obtained above we get
the following expression for decay width
\ba
d\Gamma(s_\pi,s_l,...)=\int\limits_{s_l}^{s_{max}}\frac{d s}{s}
\brs{\delta(1-\frac{s_l}{s})P_\delta+
\sigma P(\frac{s_l}{s})+\frac{\alpha}{\pi}K(\frac{s_l}{s})}d\Gamma_B(s_\pi,s,...),
\ea
with
\ba
P(x)=\br{\frac{1+x^2}{1-x}}_+=\lim_{\Delta\to 0}
\brs{\theta(1-x-\Delta)P_\theta+
\delta(1-x)P_\delta(x)},
\ea
and the quantities $P_\Delta, P_\theta(x)$ given above.
The generalized function $P(x)$ is the kernel of the evolution equation
of partonic operators of twist two \cite{Kuraev:1985hb}.

\section{Generalization to higher orders}

The obtained result  for  the decay width with the radiative
corrections in the lowest order of
perturbation theory (PT) taken into account, permits the generalization to higher
orders of PT in the so called leading logarithmic approximation (LLA).

Moreover   the terms of order $\sigma^n \alpha$  (next to leading approximation (NLO)) as
well can be taken into account if the explicit form of a $K$-factor is known.

In such a way we obtain
\ba
d\Gamma(s_\pi,s_l,...)=\int\limits_{s_l}^{s_{max}}
\frac{ds}{s} d \Gamma_B(s_\pi,s,...)
D\br{\frac{s_l}{s},\sigma}
\br{1+\frac{\alpha}{\pi} K\br{\frac{s_l}{s}}},
\ea
with the structure function  $D^{NS}(x,\sigma)=D(x,\sigma)$  has a form:
\ba
D^{NS}(x,\sigma)&=&\delta(1-x)+\sigma P(x)+\frac{1}{2!}\sigma^2 P^{(2)}(x)+...\nn\\
P^{(n)}(x)&=&\int\limits_x^1\frac{dy}{y}P\br{y}P^{(n-1)}\br{\frac{x}{y}},
\quad
P^{(1)}(x)=P(x),\nn\\
\sigma&=&\frac{\alpha}{2\pi}(2\rho-1), \quad n=2,3,...
\ea
In applications it is convenient to use the smoothed form of $D(x,\sigma)$
\ba
D(x,\sigma)=2\sigma (1-x)^{2\sigma-1}\br{1+\frac{3}{2}\sigma}-(1+x)\sigma+O(\sigma^2).
\ea
The quantity $K$ accumulates all terms which are nonsingular in the limit of
zero positron mass. It includes the contribution from emission of virtual and real photons.
Its explicit form is given in Appendix~\ref{AppendixKFactors}.
In the Table~\ref{TableKFactors} we cite the values of $K$
for several typical values of the kinematic parameters.

Using the above  expressions we can written the final expression for decay width in the form
(we imply the smooth behavior of the Born width)
\ba
\frac{d\Gamma}{d\Gamma_B}=S(\beta)\br{1+\frac{\alpha}{\pi}K}F\br{\frac{s_l}{s_{max}},\sigma}
\ea
\ba
F(z,\sigma)=2\sigma\br{1+\frac{3}{2}\sigma}\int_z^1\frac{dx}{x}(1-x)^{2\sigma-1}-
\sigma\br{\ln\frac{1}{z}+1-z}.
\label{FDef}
\ea
In experimental set-up when the averaging on the positron spectrum is accepted all
the dependence on positron mass disappears in correspondence with Kinoshita-Lee-Nauenberg
theorem
\ba
\int\limits_0^{s_{max}}\frac{d\Gamma}{d s_l}ds_l=
S(\beta)\int\limits_0^{s_{max}}\frac{d\Gamma_B}{d s_l}
\br{1+\frac{\alpha}{\pi}K}d s_l.
\ea
The vanishing of dependence on positron mass  in our case it due to the structure
function normalization  $\int_0^1 D(x,\sigma) dx=1$.

\begin{figure}[!ht]
\begin{center}
\includegraphics[width=0.9\textwidth]{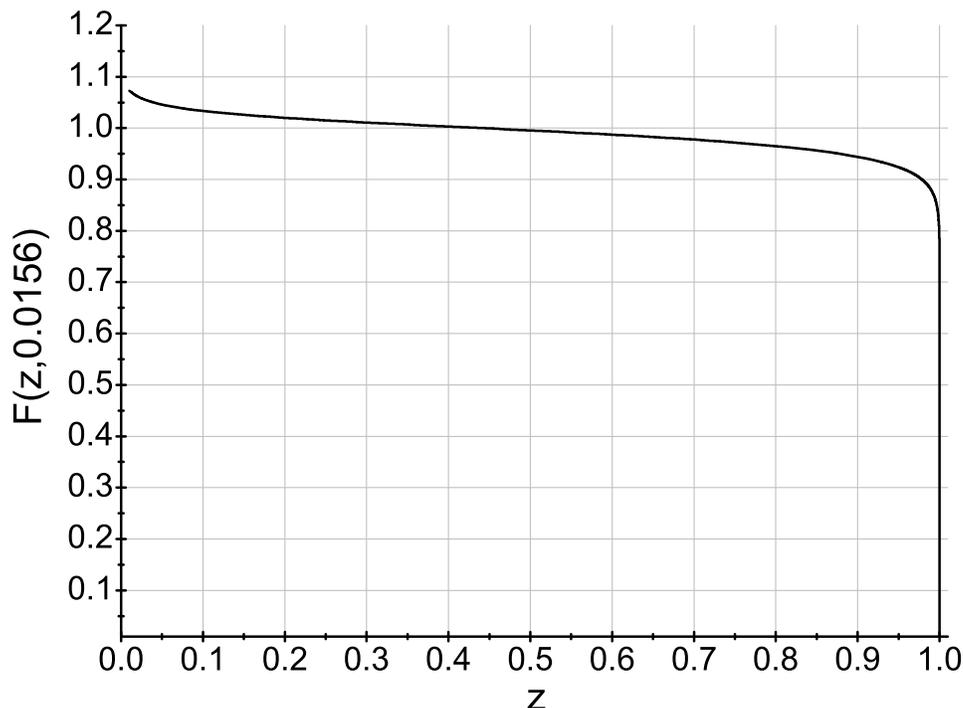}
\vspace{-1.25cm}
\caption{The dependence of fragmentation function on lepton energy (see (\ref{FDef})).
\label{FigF}}
\end{center}
\end{figure}

The values of $K$-factors for different kinematic variables
(see eqs. (\ref{DefK}), (\ref{DefKv}), (\ref{DefKs}))
and the function $F(z,\sigma)$ (eq. (\ref{FDef})) for $\sigma=0.0156$
are presented in the Table~\ref{TableKFactors} and on Fig.~\ref{FigF}.

\section{Summary}

We calculated the full set of radiation correction for the decay  width  of the
$K_{e4}$ in the lowest order in fine structure constant.
It is shown that the sum of contribution from virtual and soft real
photons emission is independent of fictitious mass of photon $\lambda$.
The ratio of the decay width to  its Born approximation is proportional to
Sommerfeld-Sakharov factor, leading to the enhancement of the
radiation correction at small relative velocity of  two charged pions in
the final state. The radiation of hard photons has been taken in account.
It has been shown that all terms including large logarithms (including
parameter $\rho=\ln{\frac{2E_e}{m_e}}$) are factorized in separate
factor which depends on the correlation between electron and pions
energies.
The utilized approach allow us to generalize the low order results to higher
orders of perturbative theory not only in leading
logarithmic approximation (LLA), but even in next to leading order
approximation (NLA). The numerical calculations are done
for K factor and fragmentation function $F(\frac{s_l}{s_{max}},\sigma)$.

\begin{acknowledgments}
We are grateful to Alexander Tarasov for interest in this problem and useful discussions.
The work of two of us (Yu.~M.~B. and E.~A.~K.) was partially supported by
INTAS grant 05-1000008-8528.
\end{acknowledgments}

\begin{table}
\begin{center}
\begin{tabular}{|c|c|c|c|c|c|c|c|}
\hline\hline
$x_+ $&$ x_-$ & $x_e$ & $y_+$ &$ y_-$ &$ K_s$&$K_v$&K \\
\hline\hline
\textbf{0.3} &\textbf{0.3} &\textbf{0.4}&\textbf{2} &\textbf{2} &\textbf{-1.99}&\textbf{-6.12}&\textbf{-5.72}\\
\hline
\textbf{0.3}&\textbf{0.3}&\textbf{0.4}&\textbf{3} &\textbf{1} &\textbf{0.83}&\textbf{-8.06}&\textbf{-5.79}\\
\hline
\textbf{0.3} &\textbf{0.3} &\textbf{0.3}&\textbf{1} &\textbf{2} &\textbf{-3.42}&\textbf{-4.45}&\textbf{-5.01}\\
\hline
\textbf{0.3}&\textbf{0.3}&\textbf{0.3}&\textbf{1.5} &\textbf{1.5} &\textbf{-1.71}&\textbf{-5.85}&\textbf{-5.30}\\
\hline
\textbf{0.3} &\textbf{0.3} &\textbf{0.2}&\textbf{0.9} &\textbf{0.9} &\textbf{-1.37}&\textbf{-5.86}&\textbf{-4.82}\\
\hline
\textbf{0.3} &\textbf{0.3} &\textbf{0.1}&\textbf{0.25} &\textbf{0.25} &\textbf{-0.91}&\textbf{-6.79}&\textbf{-4.14}\\
\hline
\textbf{0.3}&\textbf{0.4}&\textbf{0.2}&\textbf{2} &\textbf{1} &\textbf{-0.14}&\textbf{-7.54}&\textbf{-5.81}\\
\hline
\textbf{0.3}&\textbf{0.4}&\textbf{0.2}&\textbf{1.5} &\textbf{1.5} &\textbf{-1.59}&\textbf{-6.14}&\textbf{-5.26}\\
\hline
\textbf{0.4}&\textbf{0.4}&\textbf{0.1}&\textbf{1} &\textbf{2} &\textbf{-3.01}&\textbf{-6.31}&\textbf{-4.68}\\
\hline
\end{tabular}
\end{center}
\caption{\label{TableKFactors}
The values of the $K$-factors at some typical values of
kinematical variables (see eqs. (\ref{DefK}), (\ref{DefKv}), (\ref{DefKs})).
}
\end{table}

\appendix

\section{Integrals}
\label{AppendixIntegrals}

\subsection{Virtual photons emission}

Applying the Feynman denominators joining procedure and performing the loop momentum
integration, we obtain the explicit expressions  for the integrals
$I_i$ in interference term (\ref{JDef})
through the Feynman parameter
\ba
I_1&=&1+L_\Lambda-l+\int_0^1\frac{dx}{X_1}\left (2x+2(1-x)x_+-2x_+(\ln X_1+l+L_\lambda)\right);\nn\\
I_2&=&-I_1(x_+\to x_-);\quad X_1=x^2+(1-x)^2\eta^2+2x(1-x)x_+; \nn\\
I_3&=&1+L_\Lambda-l+\int\limits_0^1\frac{dx}{X_3}\left (2x+4(1-x)x_e-2x_e(\ln(\frac{X_3}{\eta^2})+L_\lambda)\right), \nn\\
X_3&=&x^2+(1-x)^2 \frac{m_e^2}{M^2} + 2x(1-x)x_e; \nn\\
I_4&=&1+L_\Lambda+\frac{2}{1-\beta^2}\int\limits_0^1
\frac{dx}{X_4}\left((1+\beta^2)(\ln{X_4}+L_\lambda)-\beta^2\right), \nn\\
X_4&=&1 - \frac{4x(1-x)}{1-\beta^2}; \nn\\
I_5&=&1+L_\Lambda+\int\limits_0^1\frac{dx}{X_5}\left(2x-2(1-x)y_--y_-(\ln{X_5}+L_\lambda)\right); \nn\\
I_6&=&-I_5(q_- \to q_+); \quad  X_5= x^2 + \frac{m_e^2}{m^2}(1-x)^2 - y_-x(1-x)].
\label{Idef}
\ea
Here we use the following notations:
\ba
x_\pm &=& \frac{E_\pm}{M};\quad  x_e = \frac{E_e}{M};\quad  y_+=\frac{2q_+p_e}{m^2};
\quad y_-=\frac{2q_-p_e}{m^2}; \nn\\
y_q&=&\frac{2q_+q_-}{m^2}=\frac{2(1+\beta^2)}{1-\beta^2};\quad  \eta = \frac{m}{M}.
\ea
Using  the neutrino shell mass condition ($p_\nu^2=0$)  one obtains  the following relation
between  these variables
\ba
y_q=\frac{2}{\eta^2}(-\frac{1}{2}+x_++x_-+x_e)-2-y_+-y_-.
\ea
The   integration  in (\ref{Idef}) can be done with the result
(we systematically omit the terms
which do not contribute in the limit of zero positron mass)
\ba
\frac{2pq_+}{M^2}\int_0^1{\frac{dx}{X_1}}&=&\frac{1}{\beta_+}L_+;\quad
\frac{2q_-q_+}{m^2}Re \int_0^1\frac{dx}{X_4}=-\frac{1+\beta^2}{\beta}L_\beta ;\nn\\
\frac{2pp_e}{M^2}\int_0^1\frac{dx}{X_3}&=&2\rho; \quad
\frac{2p_eq_-}{m^2}Re \int_0^1\frac{dx}{X_5}=-\ln {y_-}-\ln{\frac{m}{m_e}}\nn\\
\frac{2pp_e}{M^2}\int_0^1\frac{dx}{X_3}\ln X_3&=&
\ln^22x_e-\frac{1}{2}\ln^2\frac{M^2}{m_e^2}-2Li_2(1-\frac{1}{2x_e}); \nn\\
\frac{2q_+q_-}{m^2}Re\int_0^1\frac{dx}{X_4}\ln X_4&=&
\frac{1+\beta^2}{\beta}\left(\pi^2+Li_2(-\frac{2\beta}{1-\beta})-Li_2(\frac{2\beta}{1+\beta})\right);\nn\\
\frac{2p_eq_-}{m^2}Re\int_0^1\frac{dx}{X_5}\ln X_5&=&\pi^2-\ln^2 y_-
+\frac{1}{2}\ln^2\frac{m^2}{m_e^2}+2Li_2(1+\frac{1}{y_-}).
\ea

\subsection{Soft photons emission}

The integration in (\ref{SoftIntegration}) has been done using the relation
 \ba
 \int_0^{\Delta \epsilon}\frac{k^2 d k}{\omega^3}f\br{\frac{k}{\omega}}=
 \int_0^{\Delta \epsilon}\frac{d k}{\omega}f\br{\frac{k}{\omega}}-
  \int_0^{\Delta \epsilon}\frac{\lambda^2 d k}{\omega^3}f\br{\frac{k}{\omega}}
 \ea
 where $\omega=\sqrt{k^2+\lambda^2}$ with  $\lambda$  the "photon mass".
 Introducing the variable $t=k/\omega$  one obtains
 \ba
  \int_0^{\Delta \epsilon}\frac{k^2 d k}{\omega^3}f\br{\frac{k}{\omega}}=
  f\br{1}\ln\br{\frac{2\Delta \epsilon}{\lambda}}+
  \int_0^1\frac{d t}{1-t^2}\brs{t^2f\br{t}-f\br{1}}.
 \ea
Angular integration  has been done using the relation ~\cite{Kuraev:1980}
 \ba
 d\Omega=2\frac{d c_1 d c _2}{\sqrt{1-c_1^2-c_2^2-c^2+2cc_1c_2}},
 \ea
Here  $c_{1,2}$ are the cosine of the angles between 3-vectors $\vec{k}$ and $\vec{p}_{1,2}$ and
 $c$ is the cosine of the angle between the 3-vectors $\vec{p}_1$ and $\vec{p}_2$.\\
Using these relations one gets
\ba
\left.\frac{1}{4\pi}\int\frac{d^3 k}{\omega}\frac{p_1p_2}{(p_1k) (p_2k)}\right|_{\omega<\Delta \epsilon}&=&f_{12}(1)\ln(\frac{2\Delta \epsilon}{\lambda})+\int_0^1\frac{d t}{1-t^2}(t^2f_{12}(t)-f_{12}(1));\nn\\
f_{12}(t)=\frac{1}{4\pi}\int\frac{d\Omega}{(1-b_1 c_1)(1-b_2c_2)}\frac{(p_1p_2)}{\epsilon_1\epsilon_2}
 &=&\frac{1-\beta_1\beta_2 c}{\sqrt{d}}\ln\frac{1-b_1b_2c+\sqrt{d}}{\sqrt{(1-b_1^2)(1-b_2^2)}};\nn\\
\beta_1\beta_2 c=1-\frac{(p_1p_2)}{\epsilon_1\epsilon_2};\quad
b_{1,2}=t\beta_{1,2}, \quad \beta_i&=&\sqrt{1-m_i^2/\epsilon_i^2};\nn\\
d&=&(1-b_1b_2c)^2-(1-b_1^2)(1-b_2^2).
\ea
Substituting in this expression the relevant momenta we obtain the
terms determining the contribution
of soft photons emission in  considered decay rate
\ba
\left.\frac{1}{4\pi}\int\frac{d^3 k}{\omega}\frac{q_+q_-}{(q_+k) (q_-k)}\right|_{\omega<\Delta \epsilon}
&=&\frac{1+\beta^2}{2\beta}\ln(\frac{2\Delta \epsilon}{\lambda})+I_q;
 I_q=\int_0^1\frac{d t}{1-t^2}(t^2f_q(t)-f_q(1));\nn\\
\left.\frac{1}{4\pi}\int\frac{d^3 k}{\omega}\frac{p_eq_+}{(p_ek) (q_+k)}\right|_{\omega<\Delta \epsilon}
&=&\ln\frac{2p_eq_+}{m_e m}\ln(\frac{2\Delta \epsilon}{\lambda})-\frac{1}{4}(2\rho^2+\frac{\pi^2}{6})+I_{e+};\nonumber
 \ea
 \ba
 I_{e_+}&=&(1-\beta_+ c_+)\int_0^1\frac{d t}{1-t^2}[-\frac{1}{2}(\frac{t^2}{\sqrt{d_e(t)}}-\frac{1}{1-\beta_+c_+})\ln(1-t^2)\nn\\
&+&\frac{t^2}{\sqrt{d_e(t)}}\ln\frac{1-t^2\beta_+c_++\sqrt{d_e(t)}}{\sqrt{1-t^2\beta_+^2}}
-\frac{1}{1-\beta_+c_+}\ln\frac{2(1-\beta_+c_+)}{\sqrt{1-\beta_+^2}}]; \nn\\
I_q&=&\int_0^1\frac{d t}{1-t^2}(t^2f_q(t)-f_q(1)); \quad f_q(t)=\frac{1-\beta_+\beta_-c}{\sqrt{d_q(t)}}
\ln\frac{1-t^2\beta_+\beta_-c+\sqrt{d_q(t)}}{\sqrt{(1-t^2\beta_+^2)(1-t^2\beta_-^2)}};\nonumber
\ea
\ba
d_q(t)&=&(1-\beta_+\beta_-t^2 c)^2-(1-\beta_+^2t^2)(1-\beta_-^2t^2); \quad 
\beta_+\beta_-c=1-\frac{\eta^2y_q}{2x_+x_-};\nn\\
d_e(t)&=&(1-t^2\beta_+c_+)^2-(1-t^2)(1-t^2\beta_+^2);\quad \beta_+c_+=1-\frac{\eta^2y_+}{2x_+x_e}.
\ea
Finally we cited  the result of integration of the  squares of
terms in (\ref{SoftIntegration})
 \ba
\left. \frac{1}{4\pi}\int\frac{d^3 k}{\omega}(\frac{p}{p k})^2\right|_{\omega<\Delta \epsilon}
 &=&\ln\frac{2\Delta\epsilon}{\lambda}-1; \nn\\
\left. \frac{1}{4\pi}\int\frac{d^3 k}{\omega}(\frac{q_\pm}{q_\pm k})^2
\right|_{\omega<\Delta \epsilon}
 &=&\ln\frac{2\Delta\epsilon}{\lambda}-\frac{1}{\beta_\pm} L_\pm;\nn\\
\left. \frac{1}{4\pi}\int\frac{d^3 k}{\omega}\frac{p q_\pm}{(p k)(q_\pm k)}
\right|_{\omega<\Delta \epsilon}
&=& \frac{1}{2\beta_\pm}\brs{\ln\frac{2\Delta\epsilon}{\lambda}L_\pm
+ \frac{1}{2}Li_2(-\frac{2\beta_\pm}{1-\beta_\pm})-\frac{1}{2}
Li_2(\frac{2\beta_\pm}{1+\beta_\pm})}.\nn
\ea

\section{K-factors}
\label{AppendixKFactors}

As an independent set of kinematical variables we choose the five independent variables:
$x_+$, $x_-$, $x_e$, $y_+$, $y_-$.

The sum of terms independent from lepton mass
is written in the form of so called $K$-factor has the  form
\ba
K &=& K_v + K_s +\ln^2 (2x_e) - \frac{3}{2}\ln\eta+\ln(1-2\eta)+\ln 2+\frac{1}{2}\ln(2x_e)+\frac{3}{4}\nn\\
&+& \ln\br{\frac{1-2\eta}{\eta}}\brs{-4 - \frac{1}{\beta_-} L_- + \frac{1}{\beta_+} L_+
 + \frac{1+\beta^2}{\beta} L_\beta + 2\ln\frac{y_-}{y_+}};\label{DefK}\\
 K_v &=&  \int_0^1 \frac{dx}{X_1}\brs{ -x_+ \ln X_1 + x +2 (1-x)x_++2x_+\ln\eta }
+\frac{1}{2}\ln^2 y_+ -Li_2\br{1+\frac{1}{y_+}}\nn\\
&-&\frac{5}{2}\ln y_+ - 
\brs{
    x_+ \to x_-, y_+ \to y_-
}
-
\frac{5}{4}+\ln\eta-\frac{1}{2}\ln^2(2x_e)+Li_2\br{1-\frac{1}{2x_e}}-\ln(2x_e)\nn\\
&+& \frac{\beta}{2} L_\beta+\frac{1+\beta^2}{2\beta}
\brs{Li_2\br{-\frac{2\beta}{1-\beta}}-Li_2\br{\frac{2\beta}{1+\beta}}};\label{DefKv}\\
K_s &=& \frac{1}{2\beta_+}\brs{Li_2\br{-\frac{2\beta_+}{1-\beta_+}}-
Li_2\br{\frac{2\beta_+}{1+\beta_+}}}\nn\\
&-& 2I_{e+} -\brs{x_+ \to x_-}  +1-\frac{\pi^2}{6}+\frac{1}{2\beta_-}L_-+ \frac{1}{2\beta_+}L_++2I_q.
\label{DefKs}
\ea


\end{document}